\documentclass[a4paper,12pt]{article}
\usepackage[dvips]{graphics}
\usepackage{epsfig}
\usepackage{psfrag}
\usepackage[psamsfonts]{amssymb}
\usepackage{amsmath}
\usepackage{indentfirst}
\usepackage{amssymb}
\usepackage{wrapfig}
\usepackage{cite}
\usepackage{url}

\begin{document}

\title{Super-linear scaling of offsprings at criticality in branching processes}

\author{A. Saichev and D. Sornette}

\date{}
\maketitle

\begin{abstract}
For any branching process,
we demonstrate that the typical total number  $r_{\rm mp}(\nu \tau)$ of
events triggered over all generations within any sufficiently large time window $\tau$
exhibits, at criticality, a super-linear dependence  $r_{\rm mp}(\nu \tau) \sim (\nu \tau)^\gamma$
(with $\gamma >1$) on the total number $\nu \tau$ of the immigrants arriving at the Poisson rate $\nu$.
In branching processes in which immigrants (or sources) are characterized by
fertilities distributed according to an asymptotic power law tail with tail exponent
$1 < \gamma \leqslant 2$, the exponent of the super-linear law for $r_{\rm mp}(\nu \tau)$
is identical to the exponent $\gamma$ of the distribution of fertilities.
For $\gamma>2$ and for standard branching processes without
power law distribution of fertilities, $r_{\rm mp}(\nu \tau) \sim (\nu \tau)^2$.
This novel scaling law replaces and tames the divergence $\nu \tau/(1-n)$
of the mean total number ${\bar R}_t(\tau)$ of events,
as the branching ratio (defined as the average number of
triggered events of first generation per source) tends to $1$.
The derivation uses the formalism of generating probability functions. The corresponding prediction
is confirmed by numerical calculations and an heuristic derivation enlightens its underlying
mechanism. We also show that ${\bar R}_t(\tau)$ is always linear in $\nu \tau$
even at criticality ($n=1$). Our results thus illustrate the fundamental difference between
the mean total number, which is controlled by a few extremely rare realizations,
and the typical behavior represented by $r_{\rm mp}(\nu \tau)$.
\end{abstract}

\vskip 2 cm

\noindent
ETH Zurich\\
Department of Management, Technology and Economics\\
Scheuchzerstrasse 7, CH-8092 Zurich, Switzerland\\
saichev@hotmail.com and dsornette@ethz.ch

\pagebreak

\section{Introduction}

Our goal is to investigate the statistics of offsprings in a general class of branching processes,
including the standard versions as well as processes in which 
the immigrants (or sources) are characterized by
fertilities distributed according to an asymptotic power law tail with tail exponent $\gamma >1$.
Branching processes of interest are characterized by the so-called branching ratio $n$, defined
as the average number of triggered events of first generation per source.
Three regimes occur: subcritical ($n<1$), critical ($n=1$) and supercritical ($n>1$).
Here, we focus on the vicinity of the critical regime $n=1$.

To make the presentation concrete, we consider a general class of branching processes,
known as the Hawkes self-excited conditional Poisson process
\cite{Hawkes1,Hawkes2,Hawkes3,Hawkes4}, which is the simplest Poisson
process in which past events have the ability to trigger future events.
Many works have been performed to characterize the statistical and dynamical properties
of this class of models, with applications ranging from geophysical
\cite{KK,KK2,Ogata88,Ogata1,Ogata2,Ogata3,HS02,SaiSor07},
medical \cite{SorOsorio10} to financial systems,
with applications to Value-at-Risk modeling \cite{Chavezetal05}, high-frequency price processes
\cite{BauwensHautsch09}, portfolio credit risks \cite{Eymanetal10}, cascades
of corporate defaults \cite{Azizetal10}, financial contagion \cite{Aitsahaliaetal10},
and yield curve dynamics \cite{SalmonTham08}.

The paper is organized as follows. In section~2, we define mathematically the Hawkes branching model
and present briefly the formalism of generating probability functions (GPF) that is
useful for our purpose. We also define the mean
total number ${\bar R}_t(\tau)$ of events within a given time window $(t,t+\tau)$ and its main
properties. Section 3 recalls useful results based on GPF and gives the general expression
for the probability $p(r;\nu \tau)$ that the total number $R_t(\tau)$ of events
within the time window $(t,t+\tau)$ is equal to a given integer $r$ ($R_t(\tau)= r$), as
well as its asymptotic behaviors. Section 4 derives our main results by comparing
the dependence of the mean versus the mode of the distribution  $p(r;\nu \tau)$
as a function of the average number $\nu \tau$ of source events within a given window of duration $\tau$.
Section 4 contains both a generalization of the derivation of the linear scaling for the
mean total number and the derivation, both rigorous and intuitive, of the super-linear
scaling of the typical total number of events. Numerical calculations are presented, which
illustrate and confirm the analytical results. Section 5 concludes.

\section{Definitions and formalism}

\subsection{Definitions for the Hawkes self-excited conditional Poisson process}

In mathematical terms, the Hawkes model is specified via its conditional Poisson intensity (or rate)
$\lambda(t | H_t, \Theta)$, defined as the limit for small time intervals $\Delta$
of the probability that an event occurs between $t$ and
$t+\Delta$ divided by $\Delta$, given the whole past history $H_t$, which
is a function of the whole past according to
\begin{equation}
\lambda(t | H_t, \Theta) = \nu(t) +  \sum_{i | t_{i} < t} r_i h(t-t_i)~,
\label{hyjuetg2tgj}
\end{equation}
where the symbol $\Theta$ represents the set of parameters defined below.
Here, $\nu(t)$ represents the external background sources  occurring according to a Poisson process
with intensity $\nu(t)$, which may be a function of time.
The history $H_t = \{ t_i \}_{1 \leqslant i \leqslant i_t,~ t_{i_t} \leqslant t < t_{i_t+1} }$ corresponds
to all the timestamps of all events that occurred before the present time $t$.
The sum in expression (\ref{hyjuetg2tgj}) runs over all past events, spontaneous or triggered.
Therefore, the Hawkes process is non-Markovian. However, it can be mapped
exactly to a branching process, in the sense that each event can be tagged
with its direct first-generation descendants, their first-generation descendants and so on.
Aggregating all the Markov branches recovers the Hawkes process. It is important
to realize that the mapping to a branching process is probabilistic, in the sense that
a given genealogy has a probability weight and constitutes just one micro-realization
equivalent to the whole history \cite{Zhuangthesis,Zhuangetal02,Marsan07,sorutkin}.
This mapping to a branching process can be rephrased as implying that each
background event occurring at a rate $\nu(t)$ triggers, independently of each other,
its ``aftershocks'' of first generation, each of them triggering also independently
of each other their own descendants and so on, according to a
branching process statistics.

Each event $i$ that occurred at time $t_i$ has a fertility $r_i$, defined as the
expected number of events of first generation it can trigger. We assume that
the distribution of these fertilities $p_1(r)$ is such that
its statistical mean $\bar{r}$ exists. We refer to it as the branching ratio
\begin{equation}
n \equiv \bar{r} = \sum_{r=0}^\infty r \cdot p_1(r)~,
\label{hjujiuiu}
\end{equation}
which is defined as the average number of triggered events
of first generation per mother event. In the following, we will assume a specific
power law tail for $p_1(r)$, the probability that the total number of first generation
events triggered by a single given source is equal to $r$, in the form
\begin{equation}
p_1(r) \simeq \frac{\kappa}{\Gamma(-\gamma)} \cdot {1 \over r^{1+\gamma}}~ , ~~ \gamma\in(1,2)~,~~
\qquad r\gg 1 ~.
\label{jrujiujui}
\end{equation}
The condition $\gamma >1$ ensures that the mean $\bar{r} =n$ is finite.
The other bound $\gamma < 2$ corresponds to the
interesting regime where the variance of $r$ is mathematically infinite, for
which anomalous scaling occurs \cite{SHS}. This regime $\gamma < 2$ is
also the relevant one to describe earthquake triggering \cite{Helmsmallearth03}
and has been recently documented in the distributions of
commits and lines of codes per developer in open-source software projects
\cite{SorThomSai}. We will also discuss briefly how our results
extend to the case $\gamma >2$ and for pdf's $p_1(r)$ that are falling off
at large $r$'s faster than power laws, such as exponentials or even
distributions with compact support.
In expression (\ref{jrujiujui}), the scale parameter $\kappa$ obeys the inequality
\begin{equation}
0 < \gamma \kappa < n , \qquad (n\leqslant 1) ~,
\end{equation}
which ensures that
$0\leqslant p_1(0) \leqslant 1$ and $0\leqslant p_1(1) \leqslant 1$, as
seen from expression (\ref{gonepowgamdef}) below.

In the formulation of the Hawkes process (\ref{hyjuetg2tgj}), the
influence of each past event to the intensity $\lambda(t | H_t, \Theta)$
is mediated by the memory function
 or bare kernel $h(t-t_i)$, which quantifies the time-dependence
of the triggering potential of a previous event that occurred at time $t_i$ to produce
a new event at time $t$. Specifically, $r_i h(t-t_i)$ is the intensity (mean rate) of first
generation offsprings (daughters) of a mother event $i$ with average fertility $r_i$
that occurred at time $t_i$.
It follows from the time-dependent branching process theory that the intensity $r_i h(t)$ is
tied to the probability density function (pdf) $f(t)$ of the random waiting times between
the triggering event at time $t_i$ and its triggered events of
first generation at all possible future times $t > t_i$: $h(t) = f(t)$. In other words,
the parameterization chosen in
expression (\ref{hyjuetg2tgj}) is such that $h(t)$ is normalized: $\int_0^\infty h(t) dt =1$.

Equivalent to the specification of the distribution $p(r)$
of the random fertilities, we introduce the corresponding generating probability function (GPF)
of the fertilities of first generation triggered events:
\begin{equation}
G_1(z) = \sum_{r=0}^\infty  p_1(r) z^r~.
\label{yjuykuy}
\end{equation}
We assume the form
\begin{equation}
\label{gonepowgamdef}
G_1(z) = 1 - n (1-z) + \kappa (1-z)^\gamma + {\cal O}[(1-z)^2], \qquad \gamma\in(1,2) ~,
\end{equation}
where the expansion in (\ref{gonepowgamdef}) omits terms of order $(1-z)^2$ and
higher. Using the form (\ref{gonepowgamdef}),
we have $p_1(0)= 1-n+\kappa$, $p_1(1) = n-\gamma \kappa$ and
\begin{equation}
p_1(r) = \kappa \cdot \binom{r-\gamma-1}{r}
~, ~~~r>1~,
\end{equation}
whose asymptotic for large $r$ is given by expression (\ref{jrujiujui}).

In summary, the set of parameters involved in the Hawkes model, denoted by the symbol $\Theta$, includes the
branching ratio $n$ that controls the average fertility of events,
the parameters $\gamma$ and $\kappa$ of the pdf $p_1(r)$, the time scale and exponent of the
memory function $h(t)$.
The term $\nu(t)$ represents the external background sources
occurring according to a Poisson process
with intensity $\nu(t)$, which may be a function of time.
All other events are triggered by previous events, either background sources
or previously triggered events of earlier generations. Each event can in turn trigger
their own offsprings. This gives rise to the existence of
potentially many generations of events, depending on the value of the branching ratio $n$.
The regimes $n<1$, $n=1$ and $n>1$ are respectively called subcritical, critical and supercritical.

\subsection{Definitions of mean numbers of events}

The interpretation of the Hawkes process as the superposition of many independent
branching processes has very useful consequences. Recall that the branching process representation
means that one can reproduce exactly the same time series of events by considering
that each first generation event triggered by some background source triggers independently
its own first generation aftershocks (which are therefore second generation aftershocks of the
initial background source) and so on. In other words, any aftershock triggers
independently its own branching process, which is statistically identical to the branching process
generated by any background event occurring with rate $\nu(t)$.

Under these assumptions, the generating probability function (GPF) $G(z)$ of
the total number $R$ over all generations
of the events triggered by some background source satisfies  the transcendent equation
(see, for instance \cite{SHS}):
\begin{equation}\label{gfunceq}
G(z) = G_1\left[zG(z)\right] ~.
\end{equation}
This expression embodies the sum of triggered events over all branches and over all generations issued from a single
external source.

In the language of GPF, the mean number of the first generation aftershocks, or branching ratio, is given by
\begin{equation}\label{endef}
n \equiv  \bar{r}_1 = \frac{dG_1(z)}{dz}\bigg|_{z=1} ~,
\end{equation}
which is equivalent to definition (\ref{hjujiuiu}).
Together with equation \eqref{gfunceq}, one can easily show that the mean
number of triggered events from one source over all generations is equal to
\begin{equation}\label{meantotalaft}
\bar{R}_t = \frac{dG(z)}{dz}\bigg|_{z=1} = \frac{n}{1-n} .
\end{equation}
The divergence of the statistical average of the total number of triggered events as the branching ratio $n$
approach the critical value $1$ (one daughter per mother on average) is the symptom
of the transition from the subcritical to the supercritical regime occurring at the
critical value $n_c=1$ of the branching ratio. 

Let us now consider a time interval $(t,t+\tau)$ and investigate
the statistics of the number $R_t(\tau)$ of events occurring within that window. In the following,
we assume that the rate $\nu$ of exogenous background source events is constant.
Then, the mean value of the number $\bar{R}_\text{source}(\tau)$ of background sources
within that window is equal to the rate $\nu$ of its generating Poisson point process
multiplied by the window duration $\tau$: $\bar{R}_\text{source}(\tau) = \nu \cdot \tau$.
Since each source event triggers independently its own
aftershocks, with mean number given by expression \eqref{meantotalaft}, we thus obtain that
the mean rate $\nu_\text{all}$ including sources and their aftershocks over all generations  is equal to
\begin{equation}
\nu_\text{all} = \nu + \nu_\text{aftershocks} = \nu +
\nu \cdot \frac{n}{1-n} = \frac{\nu}{1-n}~ .
\end{equation}
Accordingly, the mean number $\bar{R}_t(\tau)$ of all events occurring within $(t,t+\tau)$ is equal to
\begin{equation}
\label{innerallmean}
\bar{R}_t(\tau) = \nu_\text{all} \cdot \tau =  \frac{\nu \tau}{1-n} ~.
\end{equation}
Notwithstanding the fact that
$\bar{R}_t(\tau)$ is infinite at $n=1$, the process is still stationary, if the rate of arrival
of the external sources is stationary. And it makes sense
to discuss typical realizations with bounded numbers of events
in any finite time window, as we show below. Indeed, the divergence of $\bar{R}_t(\tau)$
is controlled by rare realizations weighed by a heavy power law tail such that its
mean does not exist (exponent less than $1$).

\section{Statistics of windowed event numbers}

\subsection{Large window approximation and corresponding GPF}

We investigate the statistical properties of the number of events occurring within
an arbitrary  time interval $(t,t+\tau)$, in the limit where the window is
sufficiently large, or the memory kernel $h(t)$ is sufficiently short-range,
such that the main contribution to the number of events in $(t,t+\tau)$
comes from the source events and their aftershocks that occur within that window.
Under these conditions, without essential error, one can approximate the
number of events occurring within  $(t,t+\tau)$ that are triggered by each source in that window
by their total number of triggered aftershocks. Note that the condition
that $\tau$ should be sufficiently large, or $h(t)$ should be sufficiently short-range,
does not necessarily mean that the number of events within $(t,t+\tau)$ is large.
Indeed, if $\nu$ is sufficiently small, then the mean value given by (\ref{innerallmean})
may also be small, even in the near critical case $n\lesssim 1$
and within the large window approximation. We have previously
extensively studied the large window approximation and refer
to the corresponding articles \cite{SaiSor2006,SaiSor07,sorutkinsai}.

In mathematical language, this translates as follows.
Consider the GPF $\Theta_\text{noise}(z)$ of the number of source events within $(t,t+\tau)$.
Using the fact that these background source events follow a
Poissonian statistics, we have
\begin{equation}
\Theta_\text{noise}(z;\tau) = e^{-\nu \tau (1-z)}~ .
\label{trhy4uj}
\end{equation}
Using the replacement
\begin{equation}
z \quad \mapsto \quad z \, G(z)
\end{equation}
in the r.h.s. of expression (\ref{trhy4uj}),
where $G(z)$ is the GPF of the total number of events triggered by some
source event given by equation (\ref{meantotalaft}), we get
the expression for the GPF of the total number of events of all generations
in $(t,t+\tau)$ triggered by all sources before and within $(t,t+\tau)$:
\begin{equation}
\label{thetlwapexpr}
\Theta(z;\tau) = e^{-\nu\tau [1 - zG(z)]} ~ .
\end{equation}

\subsection{Mathematical expression of the distribution $p(r;\nu \tau)$ of event numbers}

Let us denote by $p(r;\nu \tau)$ the probability that the total number $R_t(\tau)$ of events
within the time window $(t,t+\tau)$ is equal to a given integer $r$ ($R_t(\tau)= r$).
We have previously derived the following expression \cite{SHS}
\begin{equation}
\label{probstheta}
p (r;\nu \tau) = \frac{\nu \tau}{r!} ~ \frac{d^{r-1} }{d z^{r-1}}
\left[G_1^r(z) \cdot e^{\nu \tau(z-1)} \right] \bigg|_{z=0},   \qquad r> 0~ ,
\end{equation}
which holds in our case within the large window approximation.
Expression (\ref{probstheta}) can be reformulated as the following Cauchy integral
\begin{equation}\label{pkepincaushy}
p(r;\nu \tau) = \frac{\nu \tau}{2\pi r}
\int_{-\pi}^{\pi} G_1^r \left(e^{i\phi}\right) \exp\left[ \nu \tau \left( e^{i\phi}-1\right)+i (1-r) \phi\right] d\phi~ .
\end{equation}

\subsection{Asymptotic behavior of $p(r;\nu \tau)$}

For $\nu \tau \gg 1$, the main contribution to the Cauchy integral \eqref{pkepincaushy} is
provided by values of $\phi$ close to $0$. For small $\phi$'s,
one can expand $e^{i\phi} \simeq 1+ i \phi$
and extend the two limits of the  Cauchy integral to $\pm\infty$. This yields
the asymptotic relation
\begin{equation}\label{perepcintphi}
p(r;\nu \tau) \simeq \frac{\nu \tau}{2\pi r}
\int_{-\infty}^{\infty} G_1^r(1+i\phi) e^{- i (r-\nu \tau -1)\phi} d\phi ~ .
\end{equation}
Expanding $G_1(z)$  given by expression \eqref{gonepowgamdef} for small $\phi$ reads
\begin{equation}
G_1(1+i\phi) = \left(1+i n \phi + \kappa (-i\phi)^\gamma \right)^r \simeq e^{i n r \phi +\kappa r (-i\phi)^\gamma} , \qquad r\gg 1~ .
\end{equation}
Substituting this expression in the integral in the r.h.s of expression  \eqref{perepcintphi}, we obtain
\begin{equation}
\label{perepcintphiexp}
p(r;\nu \tau) \simeq \frac{\nu \tau}{2\pi r}
\int_{-\infty}^{\infty} e^{\kappa r (-i\phi)^\gamma} e^{- i [r(1-n)-\nu \tau -1]\phi} d\phi~ .
\end{equation}

We introduce the change of integration variable $(\kappa r)^{1/\gamma} \phi = u$, which transforms
relation \eqref{perepcintphiexp} into
\begin{equation}
\label{prepsstd}
p(r;\nu \tau) \simeq \frac{\nu \tau}{2\pi r (\kappa r)^{1/\gamma}} \int_{-\infty}^{\infty} e^{(-i u)^\gamma} e^{-i x u} d u~ ,
\end{equation}
with the following definition of the auxiliary variable
\begin{equation}
x = \frac{r(1-n)-\nu \tau -1}{(\kappa r)^{1/\gamma}}~ .
\end{equation}
Expression (\ref{prepsstd}) has the form of a stable distribution. Defining the
standard stable distribution $f_\gamma(x)$ by its Fourier transform
\begin{equation}
\tilde{f}_\gamma(u) := \int_{-\infty}^\infty f_\gamma(x) e^{i u x} d x = e^{(-i u)^\gamma} ~,
\end{equation}
relation \eqref{prepsstd} transforms into
\begin{equation}\label{prefalstabrep}
p(r;\nu \tau) \simeq \frac{\nu \tau}{r (\kappa r)^{1/\gamma}} f_\gamma(x) = \frac{\nu \tau}{r (\kappa r)^{1/\gamma}}  f_\gamma\left(\frac{r(1-n)-\nu \tau -1}{(\kappa r)^{1/\gamma}} \right) .
\end{equation}

\begin{figure}
\begin{center}
  \includegraphics[width=0.85\linewidth]{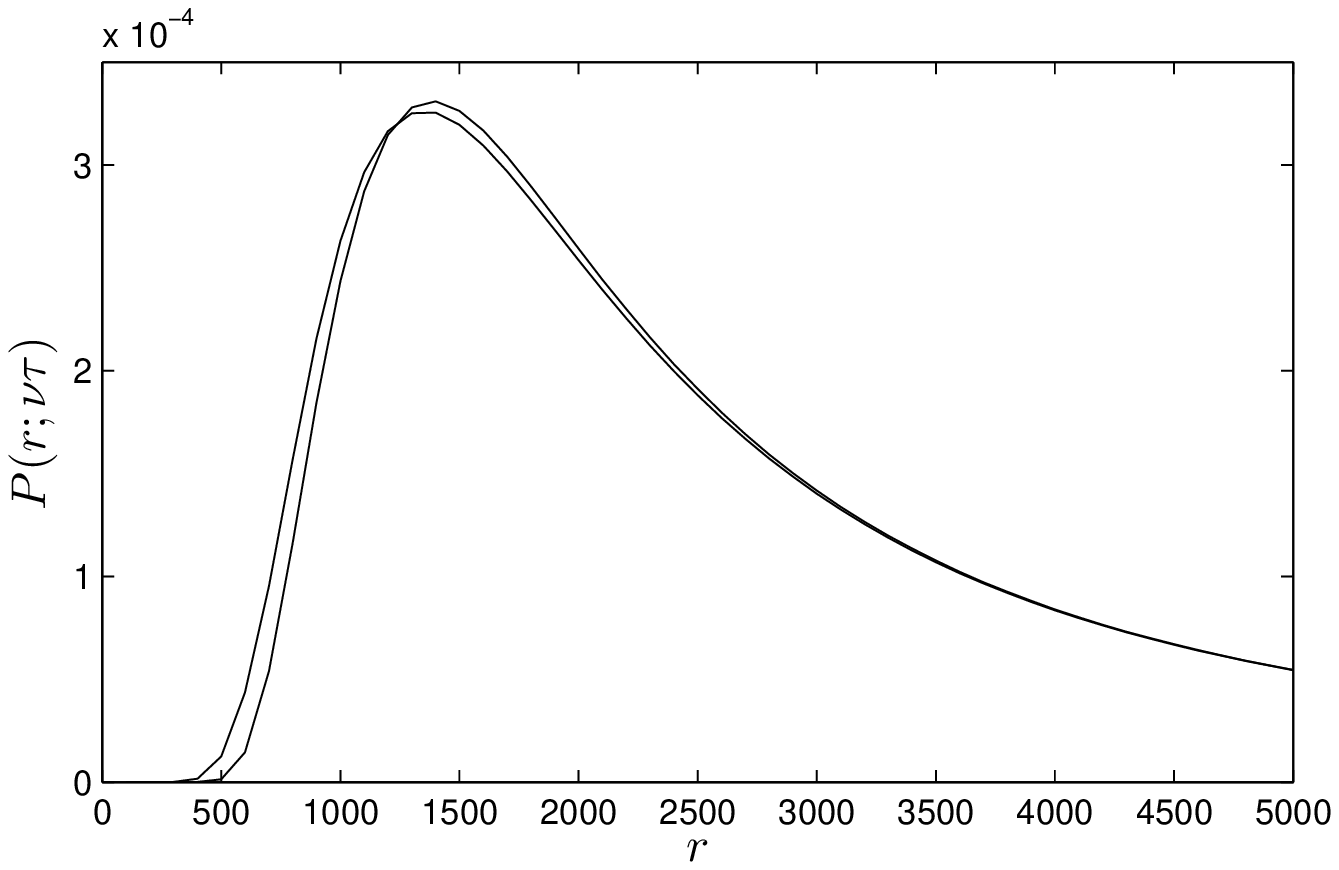}\\
\end{center}
  \caption{Plots of  the probability  $p(r;\nu \tau)$ that the total number $R_t(\tau)$ of events
within the time window $(t,t+\tau)$ is equal to a given integer $r$, given by
the ``exact'' expression \eqref{pkepincaushy} and by the asymptotic
approximation \eqref{prefalstabrep}, for $\nu \tau=100$, $n=0.99$, $\gamma=1.5$, and $\kappa=0.25$.
One can observe that the two curves differ insignificantly.}
\label{probstab}
\end{figure}

For $\nu \tau\gg 1$, the  ``exact'' expression \eqref{pkepincaushy} and its asymptotic approximation
\eqref{prefalstabrep} for the probability $p(r;\nu \tau)$ are close to each other, including their shapes
around their modes (value of  $r$ for which which $p(r;\nu \tau)$ attains its maximum), which practically coincide.
For comparison, figure~\ref{probstab} plots $p(r;\nu \tau)$ given by the two formulas
\eqref{pkepincaushy} and \eqref{prefalstabrep} as a function of $r$.

\section{Mean versus mode of the distribution $p(r;\nu \tau)$ of event numbers}

\subsection{Linear scaling of the event number mean ${\bar R}_t(\tau)$ as a function of the source rate $\nu$}

In a stationary regime where the rate $\nu(t) = \nu$ of background sources is constant,
expression (\ref{innerallmean}) shows that the mean number ${\bar R}_t(\tau)$ of all events
occurring with $(t, t+\tau)$ is simply proportional to $\nu$ and diverges as the branching
ratio approaches its critical value $n_c=1$.

We now show rigorously that the linear dependence of ${\bar R}_t(\tau)$ as a function of $\nu$ holds
true for any $\tau$'s, even when $\nu(t)$ is varying with time as for instance when there are no sources for $t<0$
and the process starts at $t=0$. In that case, expression (\ref{thetlwapexpr}) for
the GPF $\Theta(z;\tau)$ of the total number of events of all generations in $(t,t+\tau)$
triggered by all sources before and within $(t,t+\tau)$ generalizes into
\begin{equation}
\label{thetlwqer2rapexpr}
\Theta(z;\tau) = e^{-\nu \int_0^\tau [1 - zG(z,t)] dt} ~ ,
\end{equation}
where $G(z,t)$ is the GPF that generalizes $G(z)$ given by formula (\ref{gfunceq})
to account for the finite life of the process and the finiteness
of the time window $(t,t+\tau)$.
From the definition of $\Theta(z;\tau)$,  the following identity holds
\begin{equation}
{\bar R}_t(\tau) \equiv {d \Theta(z;\tau) \over d\tau}\bigg|_{z=1} ~.
\label{hhk}
\end{equation}
It follows from expression (\ref{thetlwqer2rapexpr}) that
\begin{equation}
{\bar R}_t(\tau) = \nu \tau + \nu \int_0^\tau S(t,n) dt~,
\end{equation}
where
\begin{equation}
S(t,n) \equiv {d G(z,t) \over dz}\bigg|_{z=1}
\end{equation}
is the mean number of events of all generations triggered by some source, which is counted
from the time of the occurrence of the source till time $t$. Therefore, by its definition, $S(\tau,n)$
does not depend on the rate $\nu$ of sources, which implies that the  mean number ${\bar R}_t(\tau)$ of all events
occurring within $(t, t+\tau)$ is linear in $\nu$. In the stationarity case of a constant $\nu$, this linear
dependence is expressed by relation (\ref{innerallmean}).

The divergence of ${\bar R}_t(\tau)$ at  $n \to n_c=1$ is retrieved from a different perspective.
Expression (\ref{prefalstabrep}) shows that, for $1 < \gamma < 2$,
\begin{equation}
p(r, \nu \tau) \simeq { f_\gamma(0) \over \kappa^{1/\gamma}}~
{\nu \tau \over r^{1+ {1 \over \gamma}}}~ ~{\rm for}~~ r \gg 1  ~~{\rm and}~~ n=1~.
\label{wrhrwvwrt2}
\end{equation}
Since $1 <\gamma <2$, then $1/2 < 1/\gamma <1$, which implies that ${\bar R}_t(\tau)$
is infinite for $n=1$ since it is the first-order moment of the distribution $p(r, \nu \tau)$,
which has a tail exponent $1/\gamma$ smaller than $1$. This retrieves the divergence of
relation (\ref{innerallmean}) at $n=1$.
This divergence of the mean number ${\bar R}_t(\tau)$ of all events
occurring with $(t, t+\tau)$ is thus controlled by rare realizations with exceedingly
large numbers $r$ of events. However, these rare realizations do not represent the typical
situation of a given system or of a given simulation governed by the Hawkes self-excited process.
To obtain a better understanding of the critical regime $n=1$ and of its neighborhood
($n \simeq 1$), we need to study other moments or quantiles of the distribution $p(r, \nu \tau)$
and how they behave as a function $\nu$ and $\tau$.

For $\gamma >2$ or for pdf's $p_1(r)$ falling off at large $r$'s faster than a power law,
expression (\ref{wrhrwvwrt2}) is replaced by \cite{SHS}
\begin{equation}
p(r, \nu \tau) \simeq  {1 \over r^{1+ {1 \over 2}}}~,
\label{fhujiu5u}
\end{equation}
which corresponds to the mean field distribution of cluster sizes in standard
branching processes \cite{Harris}.

\subsection{Super-linear scaling of the event number mode as a function of the source rate $\nu$}

\begin{figure}
\begin{center}
  \includegraphics[width=0.9\linewidth]{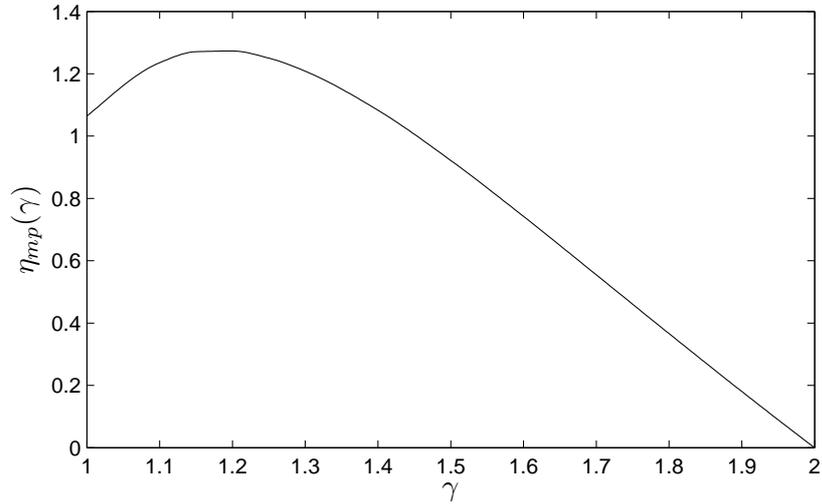}
\end{center}
  \caption{Dependence of the mode $\eta_{\rm mp}(\gamma)$ of the stable distribution $f_\gamma(x)$
  as a function of $\gamma$.}
  \label{etaplot}
\end{figure}

The principal dependence as a function of $\nu \tau$ of the mode $r_{\rm mp}$ and of the $q$-quantiles $r_q$
of $p(r;\nu \tau)$ is obtained by remarking that
the slow power factor $r^{-1-1/\gamma}$ in the r.h.s. of the asymptotic relation \eqref{prefalstabrep}
does not influence significantly their determination.
Let us call $\eta_{\rm mp}(\gamma)$ and $\eta_q(\gamma)$
respectively the mode and the $q$-quantiles of the stable distribution $f_\gamma(x)$.
Figure \ref{etaplot} shows the dependence of $\eta_{\rm mp}(\gamma)$ as a function of
$\gamma$ in the interval $1 \leqslant \gamma \leqslant 2$ of interest. Then,
the mode and quantiles of $p(r;\nu \tau)$ are given by
\begin{equation}
(1-n) r_{\rm mp} + \eta_{\rm mp}(\gamma) \left[\kappa r_{\rm mp}\right]^{1/\gamma} = \nu \tau + 1~ ,
\label{ryywww}
\end{equation}
and
\begin{equation}
(1-n) r_q + \eta_q(\gamma) \left[\kappa r_q\right]^{1/\gamma} = \nu \tau + 1~ ,
\end{equation}
From these two equations, it is obvious that the dependences of
$r_{\rm mp}$ and of $r_q$ as a function of $\nu \tau$ are identical up to a numerical constant.
We therefore only discuss the dependence of $r_{\rm mp}$.
Equation (\ref{ryywww}) yields
\begin{equation}\label{repscases}
r_{\rm mp}(\nu \tau) =
\begin{cases} \displaystyle
\frac{1}{\kappa} \left(\frac{\nu \tau}{\eta_{\rm mp}}\right)^\gamma , &
\displaystyle \nu \tau \ll  \left(\frac{\kappa \eta_{\rm mp}^\gamma}{1-n}\right)^{\frac{1}{\gamma-1}}  ,
\\[4mm] \displaystyle
\frac{\nu \tau}{1-n} ~, & \displaystyle \nu \tau \gg  \left(\frac{\kappa \eta_{\rm mp}^\gamma}{1-n}\right)^{\frac{1}{\gamma-1}} ~ .
\end{cases}
\end{equation}

The second asymptotics for
$\nu \tau \gg  \left(\frac{\kappa \eta_{\rm mp} \gamma^\gamma}{1-n}\right)^{\frac{1}{\gamma-1}}$ coincides with
the linear dependence of the mean number $\bar{R}_t(\tau)$ of events given by expression \eqref{innerallmean}.
This shows that, when $n$ is not too close to $1$, the mean number $\bar{R}_t(\tau)$ and mode are
identical asymptotically for large $\nu \tau$.

The novel insight is provided by the first asymptotics $r_{\rm mp} \sim (\nu \tau)^\gamma$, which
replaces the divergence as $n \to 1$ by a super-linear dependence with
exponent $1 < \gamma \leqslant 2$.

The two regimes in (\ref{repscases}) can be interpreted within the general framework
of bifurcations or phase transitions, where the control parameter is the branching ratio $n$,
the order parameter is the mode $r_{\rm mp}$ of the total number of events
and the external driving ``field'' is the total number $\nu \tau$ of sources.
Then, we have
\begin{equation}
r_{\rm mp} = \frac{\nu \tau}{1-n} ~, ~~~~~ 1-n \gg  {\kappa  \eta_{\rm mp}^\gamma \over (\nu \tau)^{\gamma -1}}~,
\label{heythuyjui}
\end{equation}
which expresses the standard linear relationship between order parameter $r_{\rm mp}$ and external field $\nu \tau$,
with the proportionality factor $1/(1-n)$ corresponding to the susceptibility that diverges as the critical
point $n_c=1$ is approached, with the standard mean field value exponent $1$. At criticality or very close to it,
expression (\ref{heythuyjui}) is replaced by
\begin{equation}
r_{\rm mp} \simeq \frac{1}{\kappa \eta_{\rm mp}^\gamma} \left(\nu \tau\right)^\gamma~,
 ~~~~~ 1-n \ll  {\kappa  \eta_{\rm mp}^\gamma \over (\nu \tau)^{\gamma -1}}~,
\end{equation}
which is analogous to the nonlinear dependence of the order parameter on the external
field at criticality, in the theory of critical phenomena \cite{Dombetal}.

\begin{figure}
\begin{center}
\includegraphics[width=0.9\linewidth]{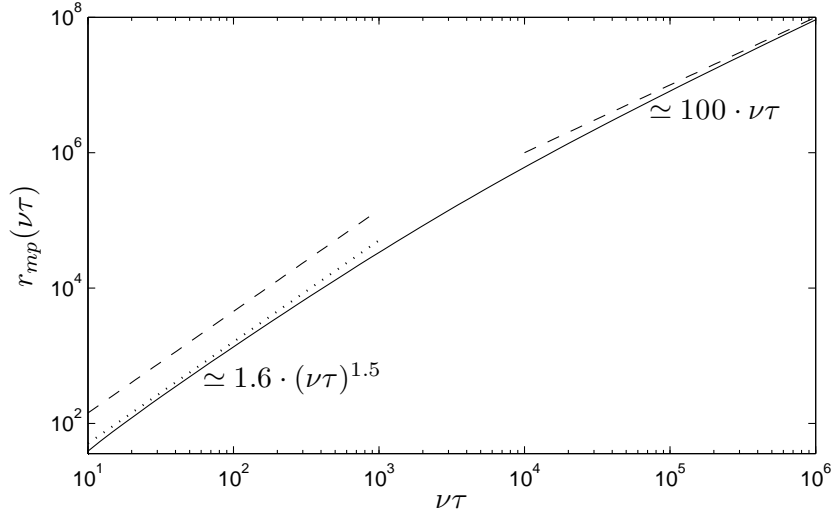}\\
\end{center}
\caption{Dependence of the mode $r_{\rm mp}(\nu \tau)$ as a function of $\nu \tau$
for $n=0.99$, $\kappa=0.25$ and $\gamma=1.5$, obtained by numerical
calculation of the integral in expression \eqref{pkepincaushy}. The dashed lines
are the asymptotics \eqref{repscases}. While the super-linear dependence
is correctly predicted, the amplitude is over estimated. The dotted line
shows the same super-linear dependence as the dashed line on the left
but with a corrected amplitude: $r_{\rm mp}(\nu \tau) \simeq 1.6 \cdot (\nu \tau)^{1.5}$.
}\label{mod150}
\end{figure}

\begin{figure}
\begin{center}
\includegraphics[width=0.85\linewidth]{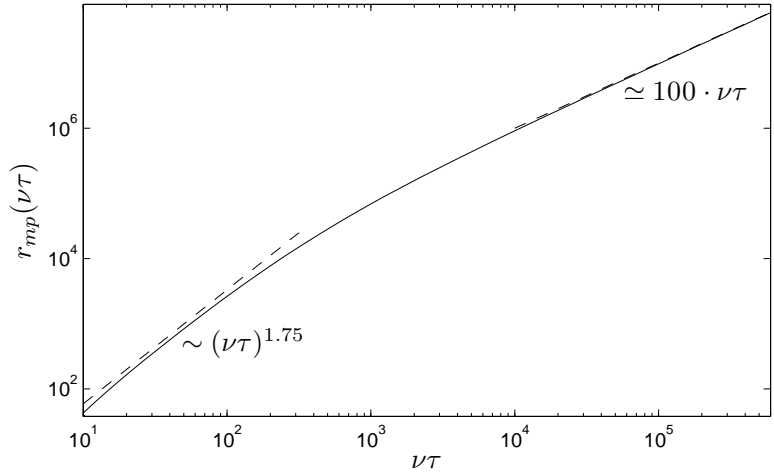}\\
\end{center}
\caption{Same as figure~\ref{mod150} for $\gamma=1.75$.}
\label{rep175}
\end{figure}

\begin{figure}
\begin{center}
\includegraphics[width=0.85\linewidth]{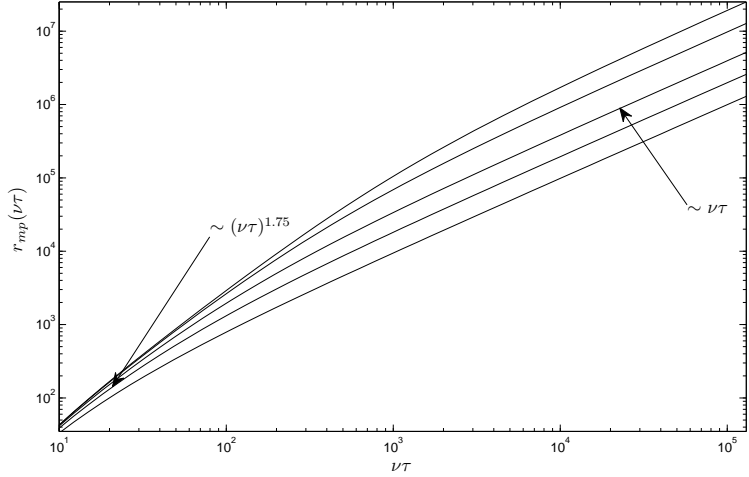}\\
\end{center}
\caption{Dependence of the mode $r_{\rm mp}(\nu \tau)$ as a function of $\nu \tau$
for $\kappa=0.25$ and $\gamma=1.75$ and different values of the branching ratio $n$
(top to bottom: $n=0.995$; $0.99$; $0.975$; $0.95$; $0.9$).
These curves are obtained by numerical calculations of the
integral in expression \eqref{pkepincaushy}.}
\label{rep175set}
\end{figure}

These two asymptotic dependences (\ref{repscases}) of $r_{\rm mp}(\nu \tau)$ can be
checked by direct numerical calculations based on expression \eqref{pkepincaushy}
for $p(r;\nu \tau)$ obtained with the large window approximation.
Figure~\ref{mod150} shows the dependence of the mode $r_{\rm mp}(\nu \tau)$
as a function of $\nu \tau$ for $n=0.99$, $\kappa=0.25$ and $\gamma=1.5$.
The two predicted super-linear and linear regimes (\ref{repscases}) are confirmed,
while the amplitude of the super-linear regime has to be corrected
according to $r(\nu \tau) \simeq 1.6 \cdot (\nu \tau)^{1.5}$.
Figure \ref{rep175} is  the same as figure~\ref{rep175set} but for $\gamma=1.75$ for which the difference
between the super-linear and linear regimes are even more striking.
Figure~\ref{rep175set} shows $r(\nu \tau)$ for $\gamma=1.75$, $\kappa=0.25$ and for
different values of the branching ratio $n$. The closer $n$ is to its critical value $1$,
the larger the range of $\nu \tau$ over which the super-linear regime
$r(\nu \tau) \sim  (\nu \tau)^{\gamma}$ holds.

\subsection{Heuristic derivation of the super-linear scaling $r_{\rm mp}(\nu \tau) \sim (\nu \tau)^{\gamma}$}

We now provide a simple intuitive derivation of the main result of this article that
the typical dependence of the total number $R_t(\tau)$ of events
within the time window $(t,t+\tau)$ as a function of $\nu \tau$ is super-linear ($\sim  (\nu \tau)^{\gamma}$
with $1 < \gamma \leqslant 2$) when the branching ratio $1$ is close to or equal to $1$.

Recall first expression (\ref{wrhrwvwrt2}) describing the tail at large $r$ of the probability  $p(r;\nu \tau)$
valid for $n=1$. Actually, the same power law tail with exponent $1/\gamma$ describes the
distribution of the total number of events of all generations generated by a given background source,
as first derived in Ref.~\cite{SHS}. As previously mentioned,
the value of the exponent $1/\gamma <1$ implies that the average number
of events of all generations per source diverges. Along the lines of reasoning developed in Refs.\cite{Bougeorges,Sornettebook04}, such mathematical divergences valid
for an infinite system translates into abnormal scaling laws for finite systems, here finite
window sizes $\tau$ and finite total number $\nu \tau$ of sources.

The reasoning goes as follows.
Within a given window of duration $\tau$, there are typically $\nu \tau$ source events, for a
Poisson arrival rate $\nu$ of the sources. Each source $i$ of these $\nu \tau$ source events produces
a random number $r_i$ of events of all generations. These random numbers
$\{r_1, r_2, ..., r_{\nu \tau-1}, r_{\nu \tau}\}$ associated with the $\nu \tau$ sources
are distributed according to $p(r) \sim 1/r^{1+{1 \over \gamma}}$ at criticality $n=1$ \cite{SHS}.
Let us call $r_{\rm max}(\nu \tau)$, the largest among this set $\{r_1, r_2, ..., r_{\nu \tau-1}, r_{\nu \tau}\}$.
A good estimate of $r_{\rm max}(\nu \tau)$ is obtained by the condition that the
probability $\int_{r_{\rm max}(\nu \tau)}^{+\infty}  p(r) dr$ to find a source with a total number
of triggered events equal to or larger than $r_{\rm max}(\nu \tau)$ times the
total number $\nu \tau$ of sources is equal to $1$. In other words,
by the definition of $r_{\rm max}(\nu \tau)$, there should be only one
source with such a total number of triggered events. This yields
\begin{equation}
r_{\rm max}(\nu \tau) \sim (\nu \tau)^\gamma ~.
\label{hruyjiu}
\end{equation}
An estimate of the typical total number of triggered events $r_1+ r_2+ ...+ r_{\nu \tau-1}+r_{\nu \tau}$
can be obtained as
\begin{equation}
r_1+ r_2+ ...+ r_{\nu \tau-1}+ r_{\nu \tau}   \approx (\nu \tau)  \int_0^{r_{\rm max}(\nu \tau)} r p(r) dr
\sim  (\nu \tau)^\gamma    ~,~~~~{\rm  for}~\gamma >1~.
 \label{rjtik6ik}
\end{equation}
Note the upper bound in the integral in estimation (\ref{rjtik6ik}) of the total
number $r_1+ r_2+ ...+ r_{\nu \tau-1}+r_{\nu \tau}$ of triggered events by all the
$\nu \tau$ sources, which reflects the fact that
the random variables $\{r_1, r_2, ..., r_{\nu \tau-1}, r_{\nu \tau}\}$ are not larger than $r_{\rm max}(\nu \tau)$
by definition of the later.
In words, equation (\ref{rjtik6ik}) states that an estimate of the typical
value of the total number of all triggered events by all sources
in the time interval $[t, t+\tau]$ is proportional to $(\nu \tau)^\gamma$, which
recovers the first scaling regime of expression (\ref{repscases}).

\begin{figure}
\begin{center}
\includegraphics[width=0.85\linewidth]{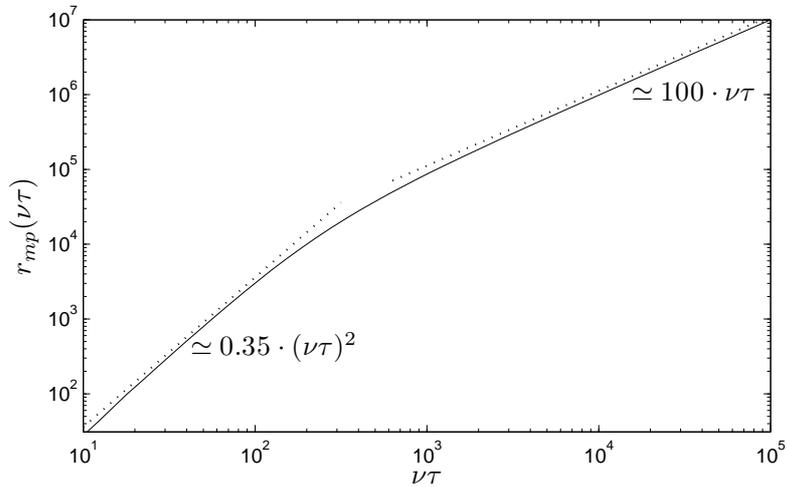}\\
\end{center}
\caption{Dependence of the mode $r_\text{mp}$ as a function of $\nu \tau$
for $n=0.99$, obtained from the exact expression (\ref{eytju46j4u}), 
corresponding to the classical Hawkes process for which $G_1(z)=e^{n(z-1)}$.
The two dotted lines show respectively the super-linear $r_\text{mp}(\nu\tau)\sim (\nu\tau)^2$ 
and linear $r\simeq 100 \cdot \nu\tau$ asymptotics.}
\label{fig6}
\end{figure}

Using expression (\ref{fhujiu5u}) and following the same line of reasoning,
the scaling relationship (\ref{rjtik6ik}) is replaced by
\begin{equation}
r_1+ r_2+ ...+ r_{\nu \tau-1}+ r_{\nu \tau}   \sim (\nu \tau)^2~,
\end{equation}
for $\gamma >2$ or for pdf's $p_1(r)$ falling off at large $r$'s faster than a power law.
This is confirmed by considering the exact derivation for the 
classical case of Poissonian statistics of first generation triggered events, for which 
the GPF of the fertilities of first generation triggered events defined by (\ref{yjuykuy})
reduces to $G_1(z)= e^{n (z-1)}$. The corresponding exact expression for 
the probability $p(r; \nu\tau)$ defined by equation (\ref{probstheta}) is
\begin{equation}
p(r;\nu \tau) = \frac{\nu \tau}{r!} \cdot \left(n r +\nu \tau \right)^{r-1} \cdot e^{-n r- \nu \tau} ~ . 
\label{eytju46j4u}
\end{equation}
Figure \ref{fig6} shows the dependence of the mode $r_\text{mp}$ of $p(r;\nu \tau)$
given by (\ref{eytju46j4u}) as a function of $\nu \tau$ for $n=0.99$,

\section{Concluding remarks}

Considering the Hawkes self-excited conditional Poisson processes, we have discovered
a novel super-linear scaling relating the typical number $r_{\rm mp}(\nu \tau) $
of triggered events by background
sources (or ``immigrants'' to employ the terminology of branching processes)
to the total number $\nu \tau$ of these sources, valid when the branching ratio is critical ($n \to 1$).
This novel scaling law $r_{\rm mp}(\nu \tau)  \sim (\nu \tau)^\gamma$ for $1 < \gamma \leq2$
and $r_{\rm mp}(\nu \tau)  \sim (\nu \tau)^2$ for $\gamma> 2$
replaces and tames the divergence $\nu \tau/(1-n)$ as $n \to 1$ of the mean total number
of triggered events. The occurrence at criticality $n=1$ of an infinite number of
generations of events, in which sources trigger daughter events, these daughters trigger
their own daughters and so on, is the mechanism for the
renormalization into a power law distribution of the total number
of triggered events over all generations per source with exponent $1/\gamma <1$
for $1 < \gamma \leq 2$ and with exponent $1/2$ for $\gamma>2$  \cite{SHS},
where $\gamma$ is the exponent of the power law distribution of
triggered events of first generation per source. In absence of such a power law
of first generation daughter per mothers, the distribution over all generations
of events generated by a single source takes the standard mean field form
of a power law tail with exponent $1/2$, as given by expression (\ref{fhujiu5u}).
Then, in the presence of this heavy tail distribution of renormalized fertilities over all
generations of the sources at criticality, the typical total number of triggered events
is dominated by the few most fertile sources in a large deviation regime, which is
responsible for the super-linear regime. We have derive this result rigorously using
the formalism of generating probability functions. The corresponding prediction
has been confirmed by numerical calculations. The essence of the mechanism
for the super-linear law has also been dissected with an heuristic derivation.

We also stress that the super-linear scale of the mode of the total number of
events over all generations within any sufficiently large time window also holds
for quantiles of the distribution $p(r;\nu \tau)$ of the total number $R_t(\tau)$ of events
within the time window $(t,t+\tau)$ that are not too far in the tails.
In contrast, we have shown that the mean of $R_t(\tau)$ is always linear in $\nu \tau$
even at criticality. Our results thus illustrate the fundamental difference between
the mean total number, which is controlled by a few extremely rare realizations,
and the typical behavior represented by $r_{\rm mp}(\nu \tau)$.

We will report elsewhere empirical evidence of the super-linear regime
in open-source software development projects \cite{SorThomSai}.


\clearpage

\end{document}